\documentclass{ptapap}

\author{Marzena \'Sniegowska}[UW,CFT]
\author{Rados{\l}aw Smolec}[CAMK]

\affil[UW]{Warsaw University Observatory, Al. Ujazdowskie 4, 00-478 Warsaw, Poland}
\affil[CFT]{Center for Theoretical Physics, Polish Academy of Sciences, Al. Lotnik\' ow 32/46, 02-668 Warsaw, Poland}
\affil[CAMK]{Nicolaus Copernicus Astronomical Center, Bartycka 18, 00-716 Warsaw, Poland}

\title{Non-radial Pulsation in the First Overtone Cepheids of the Magellanic Clouds}
\headtitle{Non-radial Pulsation in the First Overtone Cepheids}

\begin{document}

\maketitle

\begin{abstract}
We have analysed photometric data for the first overtone Cepheids in the Magellanic Clouds from the OGLE collection. In more than $500$ stars, we have detected additional variability with periods shorter than the first overtone period and period ratios in the $(0.60,\, 0.65)$ range. The sample includes double periodic stars detected previously by the OGLE team as well as new discoveries. In the Petersen diagram, these stars form three well-separated sequences. In some stars, we have found simultaneously two close periodicities corresponding to two sequences in the Petersen diagram. In a significant fraction of stars, we have detected the power excess at half the frequency of the additional variability. Interestingly, most of these stars form the middle sequence in the Petersen diagram. 
\end{abstract}

The majority of classical Cepheids are single periodic and radial pulsators, but non-radial pulsation is not rare. In particular, in many first overtone (1O) Cepheids, additional periodicities in the  $P_{\rm x}\in(0,60,\,0.65)P_{\rm 1O}$ range are detected that cannot be explained as radial modes. \cite{2008CoAst.157..343M,2009MNRAS.394.1649M} discovered such stars in the OGLE-II photometry from the Large Magellanic Cloud (LMC). \cite{2010AcA....60...17S} published a list of 138 first overtone Cepheids with additional non-radial periodicity from the Small Magellanic Cloud (SMC). These stars were analysed in detail by \cite{2016MNRAS.458.3561S}. Similar stars from the LMC were also published by \cite{2008AcA....58..163S,2015AcA....65..329S}. Here we report the initial results of our analysis of the OGLE-III and OGLE-IV \textit{I}-band photometry for 1O Cepheids from the SMC and the LMC. The goal is to find and to analyse these double-periodic variables using the complete OGLE sample.

In our analysis, we use the standard consecutive prewhitening technique. We identify the dominant frequencies with the help of the discrete Fourier transform. Then, we fit the light curve with a sine series. Next, we analyse the residuals in search for additional, low-amplitude signals. 

\begin{figure}
\centering
\includegraphics[width=0.95\hsize]{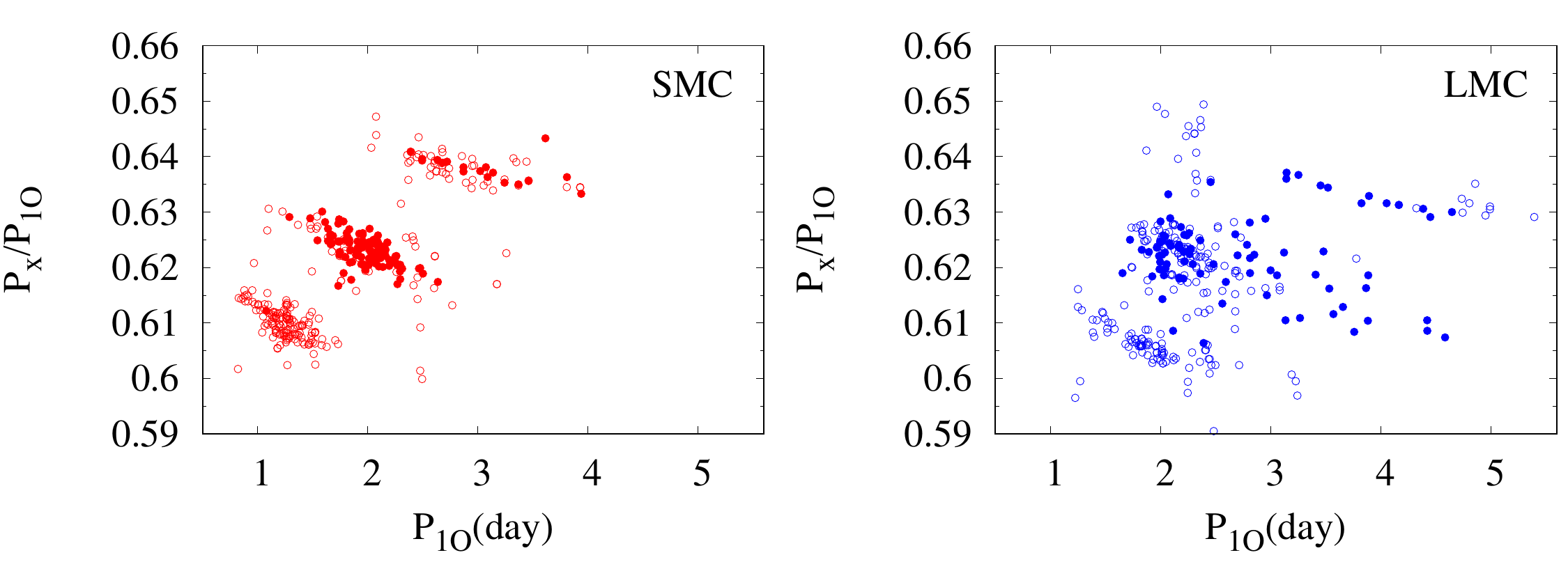}
\caption{The Petersen diagrams for classical Cepheids from the SMC (left panel) and the LMC (right panel). The filled symbols correspond to stars in which the power excess centred at $1/2\nu_{\rm x}$ was detected.}
\label{fig:pet}
\end{figure}

The Petersen diagram that shows all the double-periodic variables we have found is plotted in Fig.~\ref{fig:pet}. So far we have found 225 classical Cepheids from the LMC and 185 Cepheids from the SMC that have non-radial periodicity. Forty-one 1O Cepheids in the SMC are new discoveries, i.e., they were not reported previously in the OGLE catalogs as stars with non-radial periodicity.  The 1O Cepheids from the SMC form three well-separated sequences. The periods of the 1O Cepheids in the LMC are, on average, longer than in the SMC. A higher dispersion of the LMC stars in the Petersen diagram is noticeable. In $39$\,\% of 1O Cepheids from the SMC and in $47$\,\% of 1O Cepheids from the LMC, we detect a power excess centred at $1/2\nu_{\rm x}$ ($\nu_{\rm x}=1/P_{\rm x}$). Typically, this power excess has a complex form -- a broad cluster of peaks is detected. In the SMC, 1O Cepheids with a power excess detected at $1/2\nu_{\rm x}$ are not uniformly distributed in the Petersen diagram. The majority of these Cepheids fall within the middle sequence. In the LMC, the distribution in the Petersen diagram is similar. \cite{2016CoKon.105...23D} proposed an explanation for these double-periodic Cepheids. In his model, the signal observed at $1/2\nu_{\rm x}$ corresponds to non-radial modes of the moderate degrees, $\ell=7,\, 8$ and $9$, while the typically higher signal at $\nu_{\rm x}$ is its harmonic.

\acknowledgements{This research is supported by the National Science Center, Poland, grant agreement DEC-2015/17/B/ST9/03421.}

\bibliographystyle{ptapap}
\bibliography{ptapapdoc}

\end{document}